\documentclass{ws-procs9x6}
\begin{document}

\title{Shaping opinions in a social network}

\author{T. Carletti$^*$ and S. Righi}

\address{D\'epartement de Math\'ematique, Facult\'es Universitaires Notre-Dame
  de la Paix\\ 
Namur, Belgium, 5000\\
$^*$E-mail: timoteo.carletti@fundp.ac.be\\
www.fundp.ac.be}

\begin{abstract}

We hereby propose a model of opinion dynamics where individuals update their
beliefs because of interactions in acquaintances' group.

The model exhibit a non trivial behavior that we discuss as a function of
the main involved parameters. Results are reported on the average number of
opinion clusters and the time needed to form such clusters.

\end {abstract}


\bodymatter

\section{Introduction}
Complex Systems Science (CSS) studies the behavior of a wide range of
phenomena, from physics to social sciences passing through biology just to
mention few of them. The classical approach followed in the CSS
consists first in a decomposition of the system into \lq\lq elementary
blocks\rq\rq that will 
be successively individually analyzed in details, then the properties determined at
micro--level are transported to the macro--level. This approach results very
fruitful and shaped the CSS as an highly multidisciplinary field.

Recently models of opinion dynamics gathered a considerable amount of interest
testified by the production of specialized reviews such
as~\refcite{galam01,odbcmas,sphysocdyn}, reinforcing in this way the emergence
of the {\it sociophysics}~\cite{galam00}. A basic distinction can be done
in model of {\it continuous opinion with
  threshold}~\cite{meetandsegregate,pre}, where opinions can assumed to be
well described as continuous quantities; thus agents update their values
because of binary interactions, if their opinions are close enough, i.e. below
a given  
threshold. The second class consists of models where opinions can be described
by 
{\it discrete} variable, 
yes/no for instance, and they are updated according to {\it local rules,
  i.e. small group interactions} as
for instance: majority rule~\cite{galam02}, majority and
  inflexible rules~\cite{galam03}, majority and contrarian
  rules~\cite{galam04}. 

In this paper we introduce a new model of opinion dynamics which naturally
sets at the intersection of the former scheme; in fact individuals have
continuous opinions that are updated if they are below some given threshold,
once agents belong to a group, whose size evolves dynamically together with
the opinion. Moreover each agent possesses an {\it affinity} with respect to
any other agent, the higher is the affinity score the more trustable is the
relationship. Such affinity evolves in time because of the past interactions,
hence the acquaintances' group is determined by the underlying evolving social
network. 

We hereby provide an application of the model to the study of the
consensus--polarization transition that 
can occur in real population when people do agree on the same idea -- consensus
state -- or they divide into several opinion groups -- polarization state.

The paper is organized as follows. In the next section we introduce the model
and the basic involved parameters, then we present the results concerning the
consensus--polarization issue and the time needed to reach such asymptotic
state. We will end with some conclusions and perspectives.

\section{The model}

The model hereby studied is a generalization of the one proposed
in~\cite{pre,teowivace2008} 
because now interactions occur in many--agents groups, whose size is not fixed
a priori but evolves in time.

We are thus considering a fixed population made of $N$ agents, i.e. closed
group setting, where each agent is characterized by its opinion on a given
subject, here represented by a real number $O_i^t\in [0,1]$, and moreover each
agent possesses an affinity with respect to any other, $\alpha_{ij}^t\in
[0,1]$: the higher is $\alpha^t_{ij}$ the more affine, say trustable, the
relationships are and consequently agents behave. 

At each time step a first agent, say {\it i}, is randomly drawn with an
uniform probability, from the population; then, in order to determine its
acquaintances' group, it computes its {\it social distance} with respect to the
whole population: 
\begin{equation}
 d^t_{ij}= | O^t_i-O^t_j|\left(1-\alpha^t_{ij}\right) \, ,\forall j\in \{
 1,\dots,N\}\setminus \{ i\} \, .
\label{eq:distance}
\end{equation}
The agents {\it j} whose distance from {\it i} is lesser than a given
threshold, $\Delta g_c$, will determine the acquaintances' group of {\it i} at
time $t$ in formula:
\begin{equation}
  \label{eq:mGi}
\mathcal{F}_i^t=\Big\{j:  d^t_{ij} \leq \Delta g_c \Big\} \, .
\end{equation}
Let us observe that the group changes in time, in size and in composition,
because the opinions and/or affinities also evolve. The rationale in the use
of the affinities in the definition of the social metric is to
interpret~\cite{epjb,teowivace2008} the affinity as the adjacency matrix of
the (weighted) social network underlying the population. We hereby assume a
constant threshold $\Delta g_c$ for the whole population, but of course one
could consider non--homogeneous cases as well.  

Once the agent {\it i} has been selected and the group $\mathcal{F}_i^t$ has
been formed, the involved individuals do interact by possibly updating their
opinions and/or affinities. Once in the group, all agents
are supposed to listen to and speak to all other agents, therefore every one
can perceive a personal {\em averaged} -- by the
mutual affinity -- {\em group opinion}, $<O^t_l>$, because each agent weights
differently opinions of trustable 
agents from the others. In formula: 
\begin{equation} 
 <O^t_l> = \frac{\sum_{j=1}^{m_i} \alpha^t_{lj}  O^t_j}{\sum_{j=1}^{m_i}
   \alpha^t_{lj}} \quad \forall l \in \mathcal{F}^t_i\, , 
\end{equation}
where we denoted by $m_i$ the size of the subgroup $\mathcal{F}^t_i$. The
vector $(<O^t_1>,\dots,<O^t_{m_i}>)$, will hereby named {\it apopsicentre},
i.e. the barycentre of the opinions ($\acute{\alpha} \pi o \psi \eta$ =
opinion). 

Because the affinity is is general not symmetric, some agents could have been
included in the group determined by {\it i}, \lq\lq against\rq\rq their
advise, hence a second relevant variable is the {\em averaged affinity} that
each agent perceives of the group itself~\footnote{Each agent in the group can
  determine the apopsicentre, but if it is in an hostile group, it will not
  move toward this value.} :
\begin{equation}
<\alpha^t_l>=\frac{1}{m_i}\sum_{j=1}^{m_i} \alpha^t_{lj} \quad \forall l
\in \mathcal{F}^t_i \, .
\end{equation}

Once the former two quantities have been computed by each agent, we propose
the following update scheme: to belong in the largest size group, each agent,
would like to come closer to its perceived apopsicentre if it feels himself
affine enough to the group:
\begin{equation}
O_l^{t+1}=O_l^{t}+\frac{1}{2} \left(
  <O^t_l>-\;O^t_l\right)\Gamma_1(<\alpha^t_{l}>)  
\quad \forall l\in \mathcal{F}_i^t\, ,
\label{eq:dynopinion}
\end{equation}
where $\Gamma_1(x)=\frac{1}{2}\left[\tanh(\beta_1(x - \alpha_c))+1\right]$ is
an activating function, defining the region of trust for effective social
interactions, e.g. $<\alpha^t_{l}>$ larger than $\alpha_c$.

\begin{figure}[htbp]
\centering
\includegraphics[width=7cm]{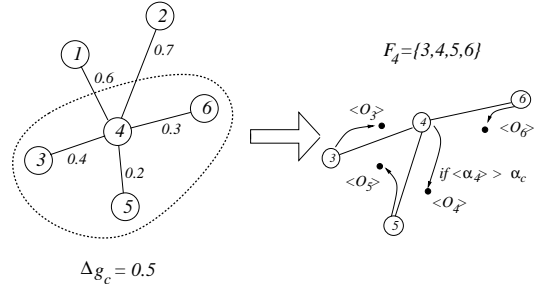}  
\caption{Cartoon to represent the group formation and the interaction
  mechanism. On the left panel, the composition of the local acquaintance
  group. On the right panel, dynamics in the opinion space : each agent tends
  to move following the represented arrows.}  
\label{fig:cartoon}
\end{figure}

Moreover sharing a close opinion, reinforce the mutual affinity, while too far
opinions make the relationship to weak, hence each agent becomes more affine
with all the agents in the subgroup that share opinions close enough to its
perceived apopsicentre, otherwise their affinities will decrease:
\begin{equation}
 \alpha^{t+1}_{jk} = \alpha^{t}_{jk}+ \alpha^{t}_{jk}
 \left(1-\alpha^t_{jk}\right) \Gamma_2(\Delta O_{jk}) \quad \forall j,k \in
 \mathcal{F}_i^t\, , 
\label{eq:dynaffinity}
\end{equation}
where $\Delta O_{jk}=<O_j^t> - \;O_k^t$, and 
\begin{equation}
 \Gamma_2(x) = \tanh\left[\beta_2 \left(\Delta O_c-|x|\right)\right] \, ,
\label{eq:gamma2}
\end{equation}
that can be considered again as an activating function for the affinity
evolution. In the previous relations for $\Gamma_{1,2}$,
we set the parameters $\beta_1$ and $\beta_2$, large enough to practically
replace the hyperbolic tangent with a simpler step function. Under these
assumptions $\Gamma_1$ takes values either 0 or 1, while the value of
$\Gamma_2$ are either -1 or 1. The interaction mechanism is schematically
represented in Fig.~\ref{fig:cartoon}. 


\section{Results}

\begin{figure}[htbp]
\centering
\includegraphics[width=7cm]{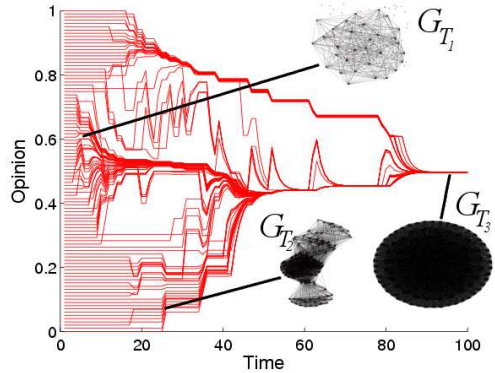}
\caption{Time evolution of the opinion (main panel) and the social network of
  affinity time-snapshots 
  (small insets). Parameters are:
  $\Delta g_c= 0.1$, $\alpha_c= 0.5$, $\Delta O_c=0.5$, $N=100$ agents whose
  opinion are 
  initially uniformly distributed in $[0,1]$, whereas initial 
  affinities are uniformly distributed in $[0,1/2]$. Three time-snapshot
  of social networks are reported for increasing times, $T_3>T_2>T_1$. Dots
  represent agents that are possibly linked if they are affine enough. Each
  network has been characterized by some standard topological indicator; the
  diameter and the averaged shortest path (that are infinite for $G_{T_1}$
  and respectively have values $4$ and $1.81$ for $G_{T_2}$ and $2$ and $1.00$
  for $G_{T_3}$), the averaged degree (that takes values $0.07$, $0.39$ and
  $0.9$ respectively for $G_{T_1}$, $G_{T_2}$ and $G_{T_3}$) and the averaged
  network clustering (that assumes the values $0.10$, $0.72$ and
  $0.99$ respectively for $G_{T_1}$, $G_{T_2}$ and $G_{T_3}$).} 
\label{figuranuovaconevolutioneimagescdelleop}
\end{figure}

A typical run of this model is presented in
Fig.~\ref{figuranuovaconevolutioneimagescdelleop}. First of we
can observe 
that the 
dynamics is faster than in the similar model presented
in~\refcite{pre,teowivace2008}, this is because binary interactions are
replaced by multi--agents interactions that improve the information
spread. Moreover there exists a transient 
interval of time, where nobody modifies its opinion, but only the mutual
affinities (see insets of
Fig.~\ref{figuranuovaconevolutioneimagescdelleop} and the relative
caption). Only when the 
relationships become highly trustable, agents do 
modify also their opinions.

This behavior is explained by the different time scales of the main
processes: evolution of opinions and evolution of affinity, as it clearly
emerges again from the insets of
Fig.~\ref{figuranuovaconevolutioneimagescdelleop}, where we show three
time-snapshots of the affinity, i.e. the social network, once agents start to
modify their  
opinions. Transferring this 
behavior to similar real social experiment, we could expect that, initially
unknown people first change (in fact construct) their affinities relationships
(increasing or decreasing mutual affinities) and only after that, they will
eventually modify their opinions. Namely initially people \lq\lq sample\rq\rq
the group and only after they modify their beliefs.

\begin{figure}[htbp]
\centering
\includegraphics[width=5cm]{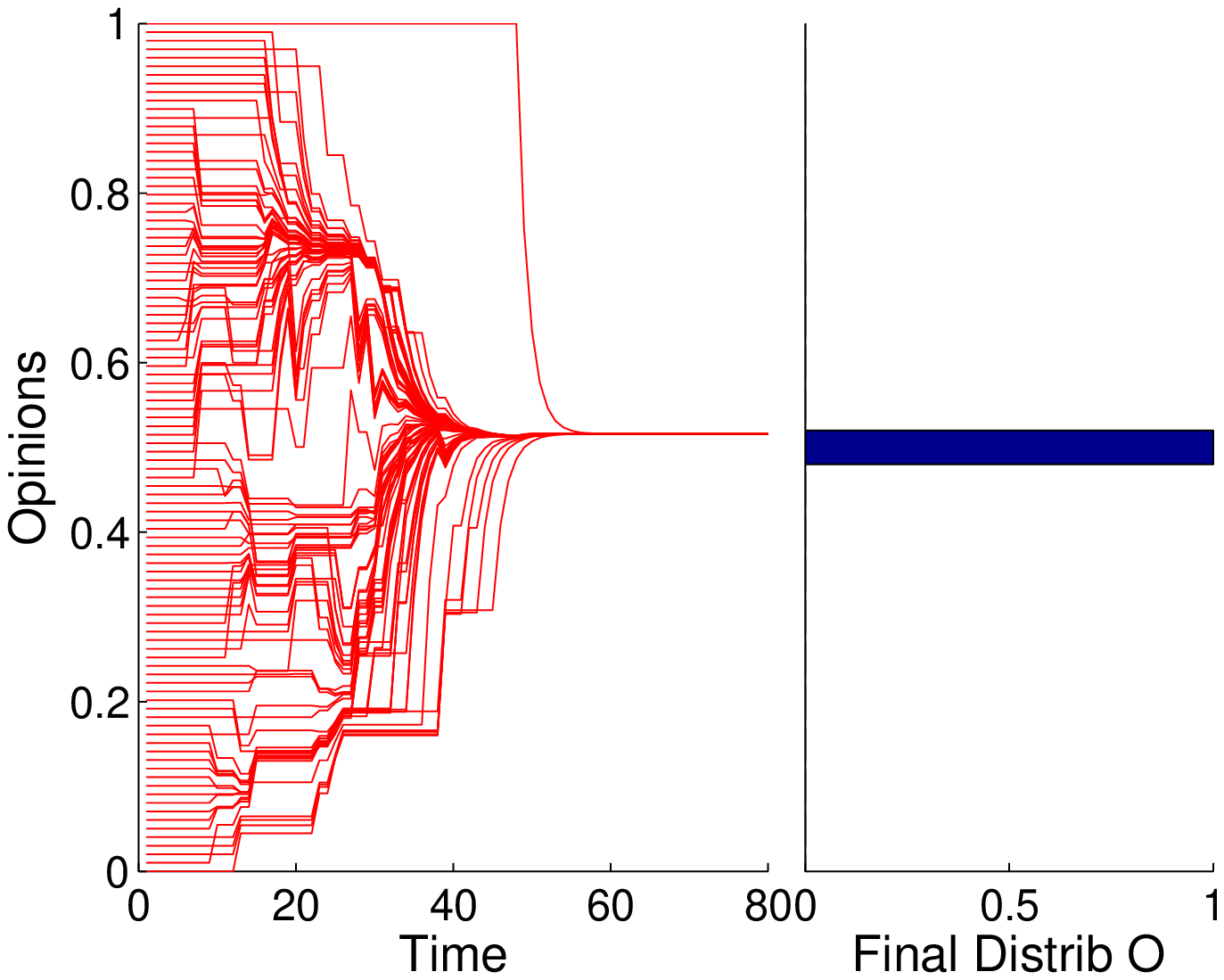}  
\includegraphics[width=5cm]{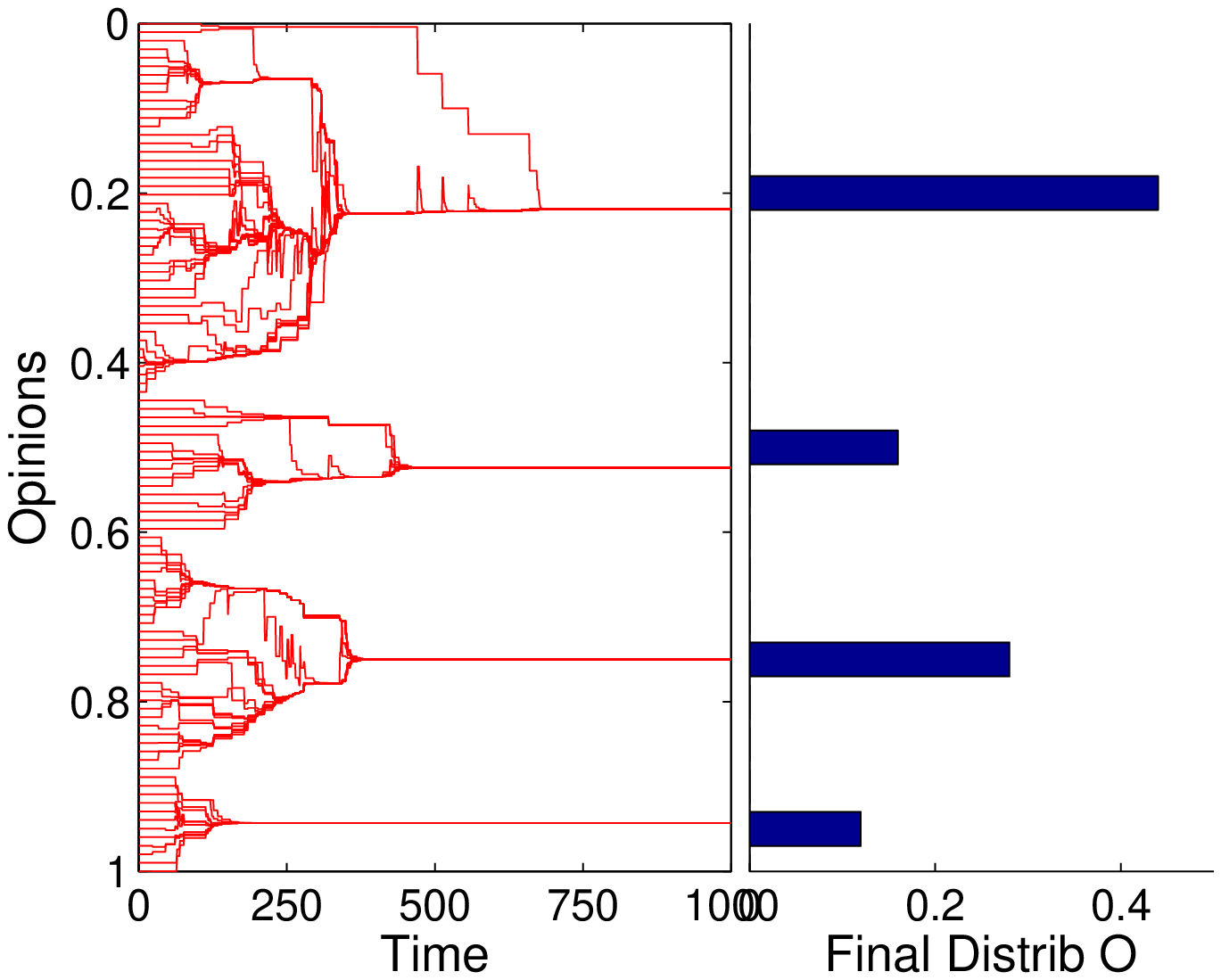}
\caption{Time evolution of the opinions and asymptotic opinion
  distribution. Parameters are: $\Delta g_c= 0.1$, $\alpha_c= 0.5$ (left
  panel), and $\Delta g_c= 0.02$, $\alpha_c= 0.5$ (right panel) both with
  $\Delta O_c=0.5$ and $N=100$ agents whose initial opinion are uniformly distributed in $[0,1]$,
  whereas initial 
  affinities are uniformly distributed in $[0,1/2]$} 
\label{figuraconconvergenzeedistrofinali}
\end{figure}

In Fig.~\ref{figuraconconvergenzeedistrofinali} we report two different
outcomes of numerical simulations of the model, for two sets of parameters, in
the left panel we found once again a consensus status, where all the population
share the same opinion, as reported by the histogram. While in the right
panel, the population polarizes~\footnote{Let us observe that polarized case
  might be metastable; in fact if the mean separation between the adjacent
  opinion peaks is smaller than the opinion interaction threshold, $\Delta
  O_c$, and $\Delta g_c$ is not too small, there always exists a finite,
  though small, probability of selecting 
  in the same acquaintance group individuals belonging to different opinion
  clusters, hence producing a gradual increase in the mutual affinities, which
  eventually lead to a merging of the, previously, separated clusters. This
  final state will be achieved on extremely long time scales, diverging with
  the group size: socially relevant dynamics are hence likely to correspond to
  the metastable regimes. A similar phenomenon has been observed
  in~\refcite{pre,teowivace2008}.} into clusters of different opinions, here
$4$. 

Hence one can characterize the final asymptotic state with the number of
opinion clusters as a function of the key parameters  $\Delta g_c$ and
$\alpha_c$. We observe that often the asymptotic state exhibits {\it
  outliers}, namely 
clusters formed by very few agents, and also that, because of the random
encounters and initial distributions, the same set of parameters can produce
asymptotic state that can exhibit a different number of opinion
clusters. For this reason we define the {\it average
  number} of opinion clusters, $<N_{clu}>$, repeating the simulation a large
number of 
times, here $500$. A second possibility is to use the {\it Deridda and
  Flyvbjerg} number~\cite{DerridaFlyvbjerg}: 
\begin{equation}
Y = \sum_{i=1}^{M} \frac{S_i^2}{N^2} \, ,
\label{eq:Y}
\end{equation}
where $M$ is the total number of clusters obtained in the asymptotic state and
$S_i$ is the number of agents in the $i$--th cluster. The quadratic dependence
on $S_i/N$ ensures that less weight has been given to small clusters with
respect to larger ones. 

\begin{figure}[htbp]
\centering
\includegraphics[width=5.5cm]{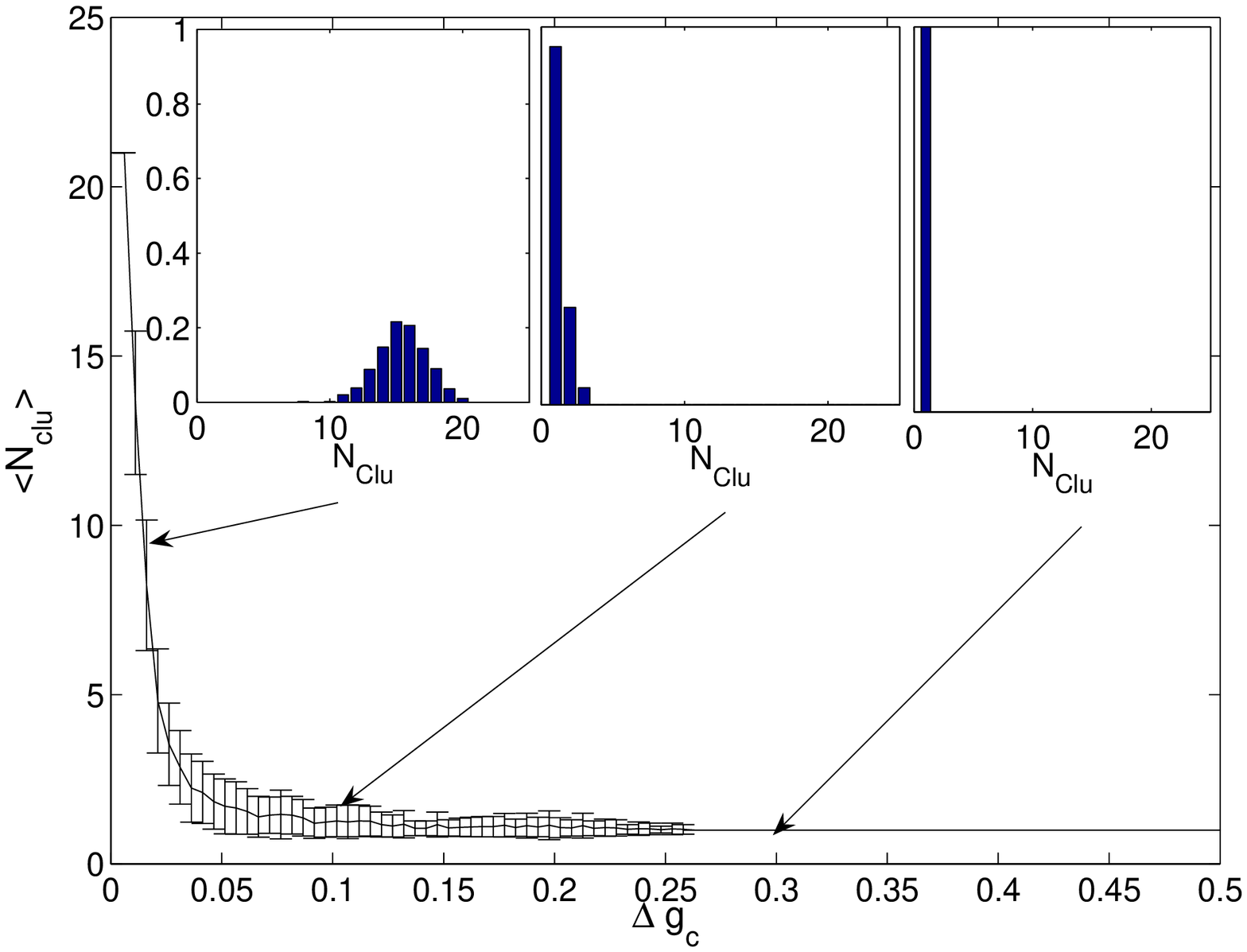}\quad
\includegraphics[width=5.5cm]{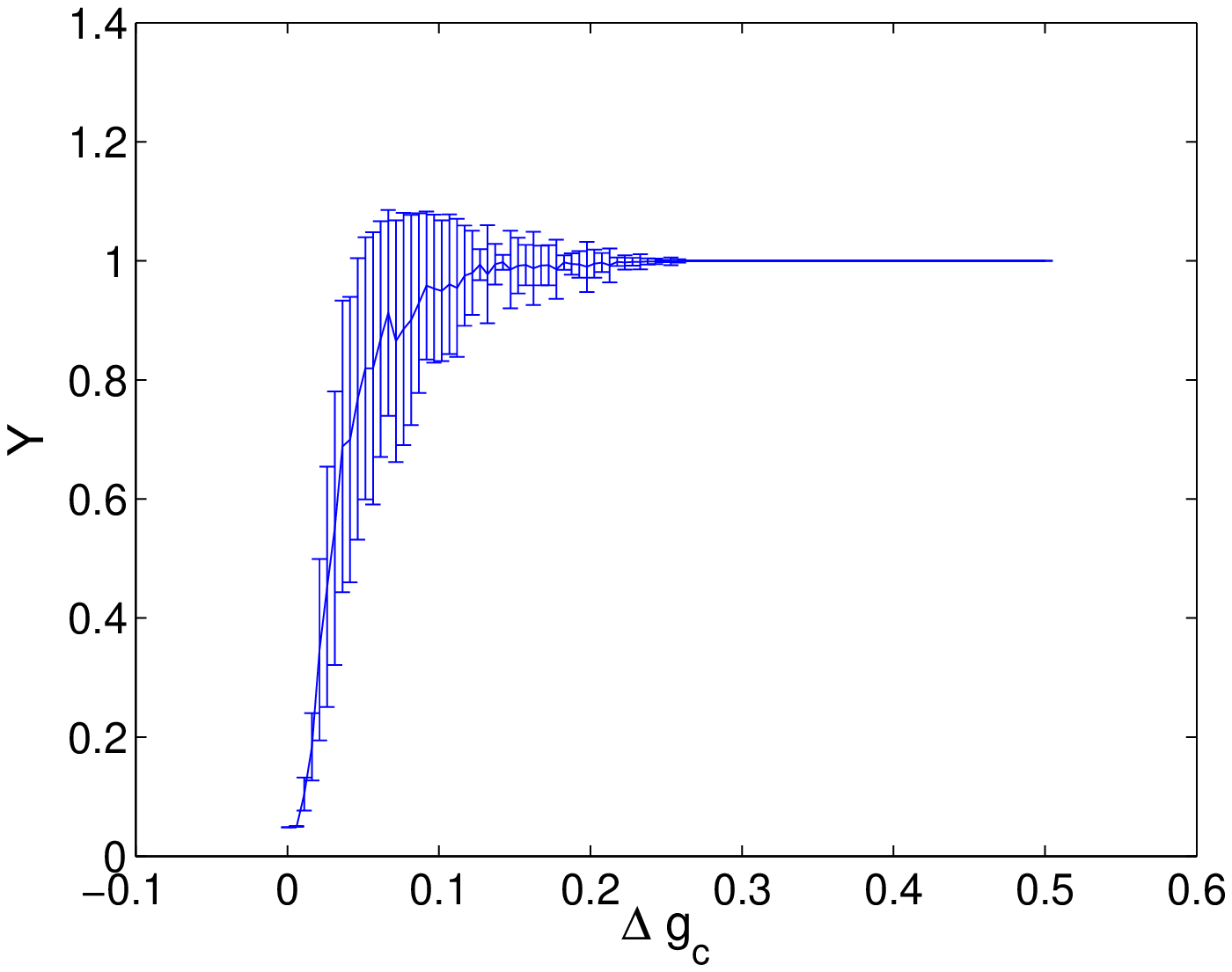}
\caption{Number of
  clusters (left panel) and Derrida and Flyvbjerg 
  (right panel)  as function of $\Delta g_c$ when $\alpha_c=0.5$. Average and
  Standard Deviation on $200$ simulations. The distributions of the
  average number of cluster are presented for $\Delta g_c=0.3$
  (right inset), $\Delta g_c= 0.1$ (central inset) and $\Delta
  g_c=0.01$ (left inset)} 
\label{fig:clumedioederiddavsdeltagc}
\end{figure}

In Fig.~\ref{fig:clumedioederiddavsdeltagc} we report the results of the
analysis of $<N_{clu}>$ and $Y$ as a function of $\Delta g_c$, for a fixed
value of $\alpha_c$.   

A phase transition from a mono--cluster state, i.e. {\it consensus}, to 
polarization of the opinions in a population of $100$ agents, emerges close to
 $\Delta g_c = 0.25$, for 
smaller values of $\Delta g_c$ the distribution of the number
of cluster can be well described by a normal distribution (see left inset 
Fig.~\ref{fig:clumedioederiddavsdeltagc}), for larger value of $\Delta g_c$,
only one  
cluster is present (see right inset Fig.~\ref{fig:clumedioederiddavsdeltagc}), while
for $\Delta g_c$ varying around $0.25$ an exponential distribution 
can be found (see middle inset Fig.~\ref{fig:clumedioederiddavsdeltagc}),
reinforcing 
thus the statement of the existence of a phase 
transition.  

\begin{figure}[htbp]
\centering
\includegraphics[width=6.5cm]{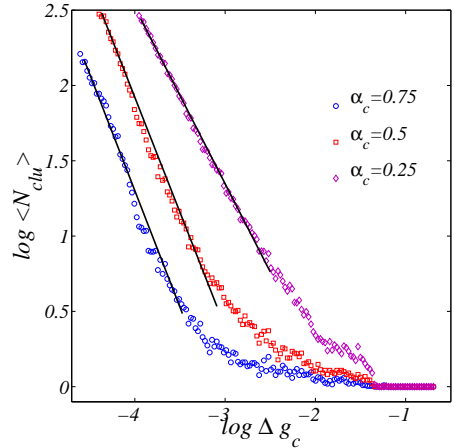}
\caption{Average number of clusters as function of $\Delta g_c$ (log--log
  scale). Best linear fits for different values of $\alpha_c$.  Each  
  simulation has been repeated $200$ times}
\label{fig:logFig7a}
\end{figure}

Data from Fig.~\ref{fig:clumedioederiddavsdeltagc} suggest an exponential
growth of $<N_{clu}>$ as a function of $\Delta g_c$ below the phase transition
value, we thus compute a linear fit on log--log scale (see
Fig.~\ref{fig:logFig7a}) in the region involving small values of 
$\Delta g_c$, obtaining:
\begin{equation*}
  \log <N_{clu}> = -1.495 \log (\Delta g_c) -4.107\, ,
\end{equation*}
when $\alpha_c=0.5$. A similar power law behavior is still valid also for the
Derrida and Flyvbjerg number. The existence of a power low seems robust with
respect to variations of the parameter $\alpha_c$ (see Fig.~\ref{fig:logFig7a}).
The results presented in Fig.~\ref{fig:logFig7a} allow us to extract also the
behavior of the average number of clusters as a function of the second
parameter $\alpha_c$ for a fixed $\Delta g_c$. In fact moving upward on vertical
lines, i.e. decreasing $\alpha_c$, the $<N_{clu}>$ increases if $\Delta g_c$
is below the critical threshold, while above this value the number of clusters
is always equal to one. Moreover from these data we could conclude that the
phase transition point seems to be independent from the value of $\alpha_c$.

Another relevant quantity of interest, is the time needed to form an opinion
cluster, the {\em opinion convergence time}, $T_c$, and its dependence of the
size of the 
cluster. Numerical simulations not reported here, emphasize that in the
polarization case $T_c$ depends in a highly
non--trivial way on the total 
number of clusters and on their sizes, roughly speaking if in
consensus state the time needed to form a cluster of say $N_1$ individual is 
some value $T_1$, then the time needed to form a cluster of the same size in a
polarization case, has nothing to do with $T_1$ and it depends on all the
formed clusters.

Because this paper offers a preliminary analysis, we decided to consider only
the consensus case, hence choosing
parameters ensuring the existence of only one 
cluster and we define the {\em convergence time} $T_c$ to be: 
\begin{equation}
T_c = \min\Big\{ t\geq 0 :  \max_i(O_i(t))-\min_i(O_i(t)) \le
\frac{a}{N}\Big\} \, , 
\label{eq:a}
\end{equation}
where $a$ is a small parameter (hereby $a=0.1$).

\begin{figure}[htbp]
\begin{center}
\mbox{\includegraphics[width=7.5cm]{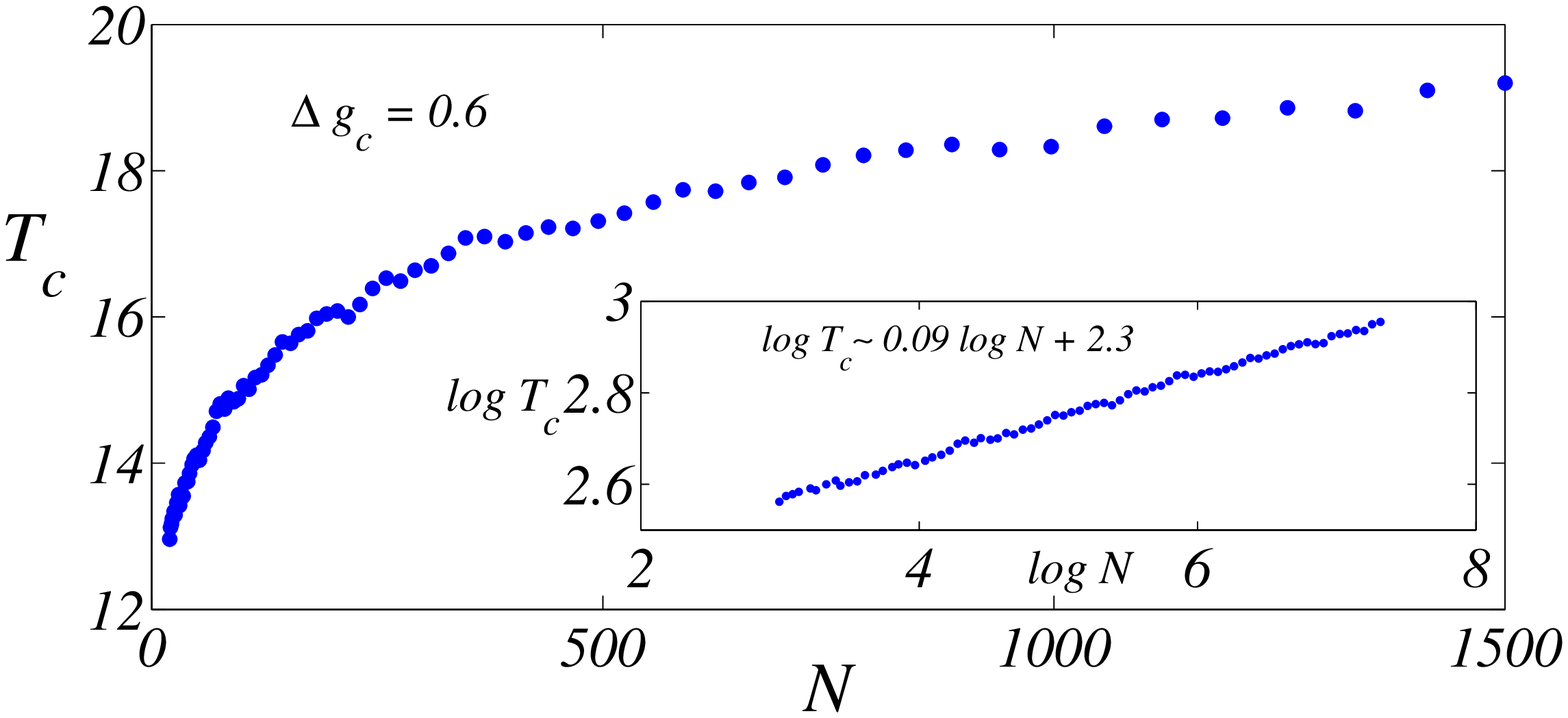}\hfill 
  \includegraphics[width=7.5cm]{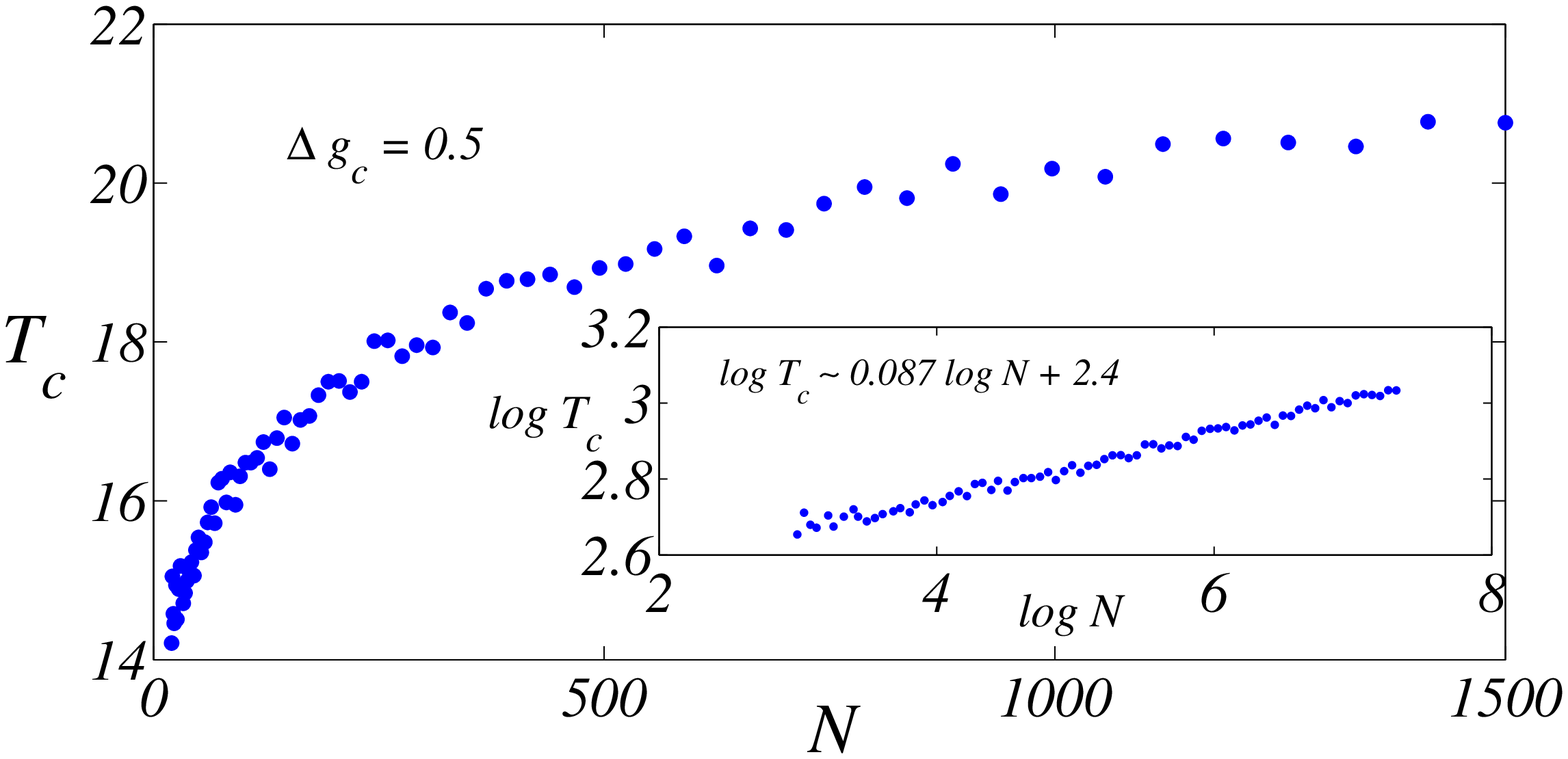}}
\includegraphics[width=7.5cm]{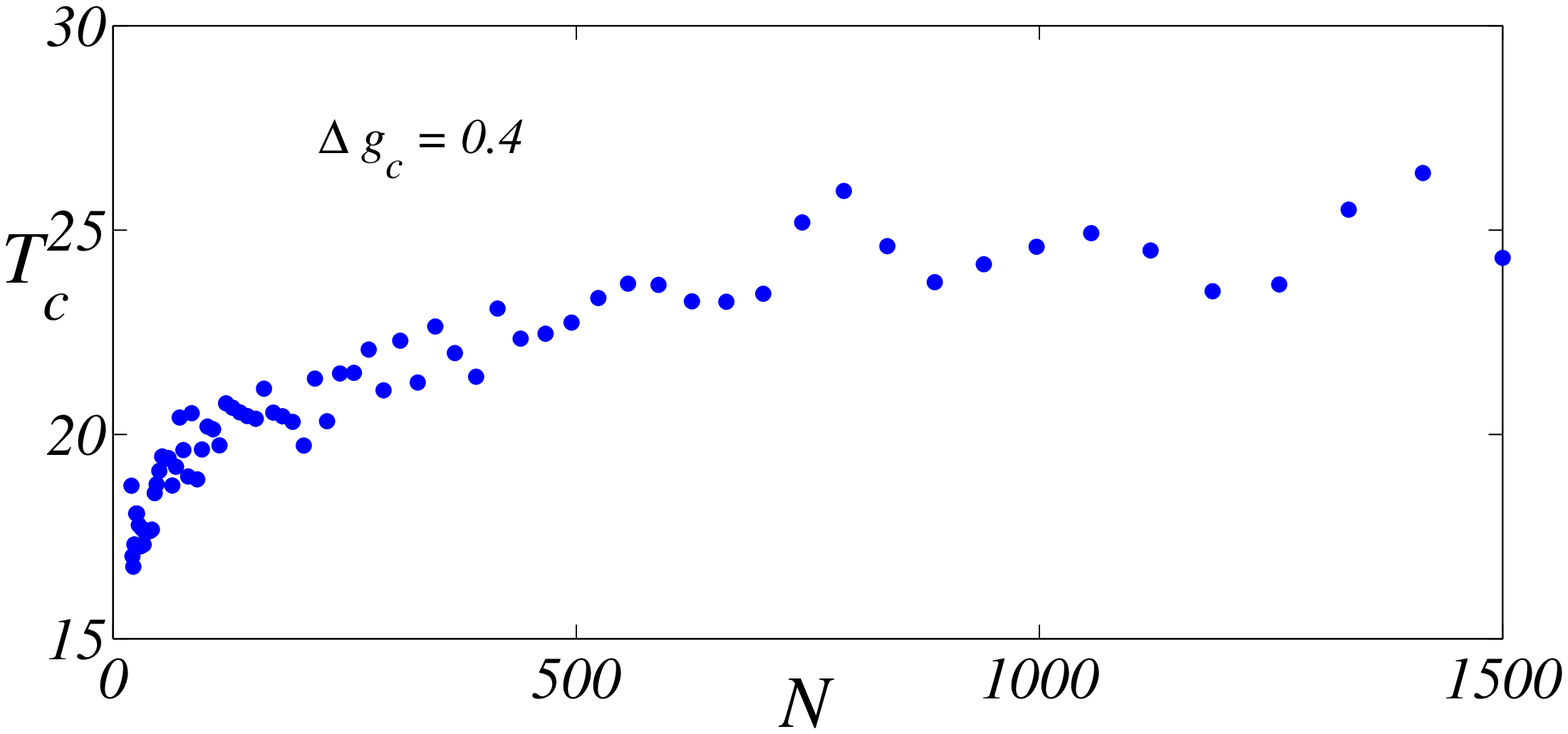}
\end{center}
\caption{Time of convergence of the opinions as function of number of
  agents. Panels correspond to parameters $\Delta g_c = 0.6$ (top-left),
  $\Delta g_c = 0.5$ (top-right) and $\Delta g_c= 0.4$ (bottom). Insets
  log--log plots of $T_c$ as a function of $N$.} 
\label{fig:opconvdist}
\end{figure}

We thus performed dedicated simulations with $N$ ranging from few unities to
thousand unities. The results reported in Fig.~\ref{fig:opconvdist} suggest a
non--linear dependence of $T_c$ on $N$, well approximable by $T_c \sim
N^b$. Using a regression analysis on the data, we can estimate the exponent
which results, $b=0.091$ for $\Delta g_c=0.6$ and $b=0.087$ for $\Delta
g_c=0.5$. Let us observe that as $\Delta g_c$ approaches the phase transition
value, $T_c$ increases and the curve becomes more noisy (see for instance the
bottom panel of Fig.~\ref{fig:opconvdist} corresponding to $\Delta g_c =0.4$),
that is because the occurrence of the consensus case becomes lesser and lesser
probable. 

\section {Conclusions}

In this paper we introduced a new model of opinion dynamics where agents
meet in social groups, affinity driven, and possibly update their beliefs
as a consequence of these local, i.e. group level, interactions. The model
exhibits a rich phenomenology determined by the interlay between the dynamics 
of the opinions and the mutual affinities. We emphasized the role
of two parameters, $\Delta g_c$ and $\alpha_c$, 
which can be qualitatively interpreted respectively as the openness of 
mind in the formation of the group and as the
openness of mind in the intra--group dynamics.
We thus studied the behavior
of the model as a function of these two parameters.

The formulation of our model has been inspired by the observation of the way
in which the formation mechanisms for social interactions do occur in the real
world: a large majority of the processes of 
formation and evolution of the opinions are driven by group based discussions,
such groups are determined by the mutual affinity and/or the shared
opinion.
The processes of group formation which tends to
form clusters of acquaintances (or collaborators) are introduced in our model
via the selection mechanism based on the mutual
trust, i.e. Eq.~\eqref{eq:distance} and~\eqref{eq:mGi}.

The numerical analysis we performed, shows a dependence of the
consensus/polarization state on the degree of mind openness in the creation of
the groups of acquaintances, i.e. the parameter $\Delta g_c$, large values
corresponding to consensus states while small ones to fragmented
groups. That is the main reason why the model exhibits a 
phase transition with respect to this variable. 
Finally the intra--group dynamics, based on the mutual affinity, allows
to update the opinions only for agents that perceive the group discussion
close enough to its believes. This phenomenon is modeled by
Eq.~\eqref{eq:dynopinion} and~\eqref{eq:dynaffinity}. 
Our analysis shows that the stronger is the degree of intra--group affinity
required to make an interaction effective, i.e. large $\alpha_c$, the higher
will be the 
degree of polarization of the population.

We can thus conclude that the model here presented, exhibits and well
reproduces the two underlying dynamical mechanisms that can drive the opinion
formation process in (relatively) small 
groups: exchange of information and mutual trust. Moreover these mechanisms
evolve on different times scales as clearly showed previously.

This model represents thus a framework where to study groups 
interactions with applications to real social systems. It would be 
interesting to improve the model by introducing, at least, two factors:
vectorial opinions \cite{axelrod, sphysocdyn, Fortunatoec} 
i.e. agents discuss and exchange information about more than one
subject. Second, introduce a limitation in the number of agents with which
anyone can be affine with, as usually is the case in the real social networks.

\end{document}